\documentclass[useAMS,usegraphicx,usenatbib]{mn2e}
\usepackage{amsmath}
\newcommand{\ppmap}{{\scriptsize PPMAP}}

\title[Multi-temperature mapping of dust structures]{Multi-temperature mapping of dust structures throughout the Galactic Plane using the PPMAP tool with Herschel Hi-GAL data}
\author[K. A. Marsh et al.]{K. A. Marsh$^{1}$\thanks{E-mail:
Ken.Marsh@astro.cf.ac.uk}, 
A. P. Whitworth${^1}$, 
O. Lomax${^1}$, 
S. E. Ragan$^{1}$,
U. Becciani$^{2}$,
\newauthor L. Cambr\'esy$^{3}$,
A. Di Giorgio$^{4}$,
D. Eden$^{5}$,
D. Elia$^{4}$,
P. Kacsuk$^{6}$, 
S. Molinari$^{4}$,
\newauthor P. Palmeirim$^{7}$,
S. Pezzuto$^{4}$,
N. Schneider$^{8,9}$,
E. Sciacca$^{2}$,
F. Vitello$^{2}$
 \\ \\
$^{1}$School of Physics and Astronomy, Cardiff University, Cardiff CF24 3AA, UK\\
$^{2}$INAF - Astrophysical Observatory of Catania, Italy \\
$^{3}$Observatoire astronomique de Strasbourg, Universit\'e de Strasbourg, CNRS, UMR 7550, 11 rue de l’Universit\'e, F-67000 Strasbourg, France \\
$^{4}$Istituto di Astrofisica e Planetologia Spaziali - INAF, Via Fosso del Cavaliere 100, I-00133 Roma Italy \\
$^{5}$Astrophysics Research Institute, Liverpool John Moores University, IC2, Liverpool Science Park, 146 Brownlow Hill, Liverpool, L3 5RF, UK \\
$^{6}$MTA Sztaki, 1111 Budapest, Kende u.\ 13-17, Hungary\\
$^{7}$Laboratoire d'Astrophysique de Marseille, CNRS/INSU-Universit\'e de
Provence, 13388 Marseille cedex 13, France \\
$^{8}$Universit\'e Bordeaux, LAB, UMR 5804, 33270 Floirac, France \\
$^{9}$CNRS, LAB, UMR 5804, 33270 Floirac, France}
\begin{document}

\pagerange{\pageref{firstpage}--\pageref{lastpage}} \pubyear{2017}

\maketitle

\label{firstpage}

\begin{abstract}
We describe new Hi-GAL based maps of the entire Galactic Plane, obtained using
continuum data in the wavelength range 70--500 $\mu$m. These maps are 
derived with the \ppmap\ procedure, and therefore represent a significant 
improvement over those obtained with standard analysis techniques. 
Specifically they have greatly improved resolution (12 arcsec)
{\it and\/}, in addition to more accurate integrated column densities and
mean dust temperatures, they give temperature-differential column densities,
i.e., separate column density maps in twelve distinct dust temperature
intervals, along with the corresponding uncertainty maps. The complete set
of maps is available online. We briefly describe \ppmap\ and 
present some illustrative examples of the results. These include
(a) multi-temperature maps of the Galactic HII region W5-E, (b) the
temperature decomposition of molecular cloud column-density probability
distribution functions, and (c) the global variation of mean dust temperature
as a function of Galactocentric distance. Amongst our findings are: (i)
a strong localised temperature gradient in W5-E in a direction orthogonal
to that towards the ionising star, suggesting an alternative
heating source and providing possible guidance
for models of the formation of the bubble complex, and (ii) the overall radial 
profile of dust temperature in the Galaxy shows a monotonic 
decrease, broadly consistent both with models of the interstellar 
radiation field and with previous estimates at lower
resolution. However, we also find a central temperature plateau
within $\sim6$ kpc of the Galactic centre, outside of which is
a pronounced steepening of the radial profile. This behaviour
may reflect the greater proportion of molecular (as opposed to atomic)
gas in the central region of the Galaxy.
\end{abstract}


\section{Introduction}
Mapping dusty structures in the Galaxy can provide key information
constraining the ecological cycle associated with star formation.
The \ppmap\ procedure \citep{mar15} is designed to extract information
from such maps in a way which takes full advantage of the observations
and knowledge of the instrumental response. Here we demonstrate the capabilities
of the method by applying it to data from the Hi-GAL survey, which used the 
{\it Herschel\/} Space Observatory to map  dust continuum emission
from the entire Galactic Plane at far-infrared wavelengths \citep{mol2010}. 
\ppmap\ represents a significant
step beyond conventional approaches to column density mapping
in which it is assumed that the dust temperature is uniform everywhere along
the line of sight \citep[see, for example,][]{kon10,per10,bern10}, 
and the observed images are smoothed to a common spatial resolution. 
In the case of {\it Herschel\/} data, this usually corresponds to the 
36 arcsec resolution at 500 $\mu$m, although some of the structure at the
18 arcsec resolution of the 250 $\mu$m data can be restored via spatial 
filtering
techniques \citep{palm13}. In contrast, \ppmap\ does not require smoothing
of the input images since it takes full account of the point 
spread functions (PSFs) of the telescopes used, and this enables
a spatial resolution of 12 arcsec to be obtained with the Hi-GAL data. 
In addition, \ppmap\ adds a third dimension to the mapping procedure, i.e.,
dust temperature. A full mathematical 
description of the procedure is given in \citet{mar15} and the essential
points are summarised in the next section. All of the Hi-GAL
data have been processed by \ppmap, and the results are available 
online\footnote{http://www.astro.cardiff.ac.uk/research/ViaLactea/}.
Here we present some preliminary results
to illustrate the type of information that can be derived.

\section{The PPMAP algorithm}

The essential concept of \ppmap\ is the {\it point process\/}, whereby the 
system under study is represented by a set of points in a suitably defined
state space. In the present context, the system is a dusty astrophysical
structure such as a core, filament or molecular cloud. It is represented by
a set of very small building-block components, each of unit column density, and
each of which is parametrised by three variables, namely the $x,y$ angular
position on the sky and dust temperature, $T_{\rm D}$. Given a set of
observational images of dust emission at multiple wavelengths, and assuming the 
dust to be optically thin, the algorithm uses a non-hierarchical Bayesian 
procedure to generate a density function 
representing the expectation number of components per unit volume of state 
space. Because of the way that a component is defined, this density function 
is equivalent to differential column density as a function
of $x$, $y$, and $T_{\rm D}$. 

The measurement model upon which the estimation is based is of the form:
\begin{equation}
{\mathbf d} = {\mathbf A}{\mathbf\Gamma} + {\mathbf\mu}\,.
\label{eq1}
\end{equation}
where ${\mathbf d}$ is the measurement vector whose $m^{\rm th}$ component 
represents the pixel value at location $(X_m,Y_m)$ in the observed image at 
wavelength $\lambda_m$; 
${\mathbf \Gamma}$ is a vector whose components represent the actual 
number of components in each cell of the state space;
$\;{\mathbf\mu}$ is the measurement noise
assumed to be Gaussian; $\;{\mathbf A}$ is the system 
response matrix whose $mn^{\rm th}$ element expresses the response of the $m^{\rm th}$ measurement to a source component occupying the $n^{\rm th}$ cell in the state space, corresponding to spatial location $(x_n,y_n)$ and dust temperature $T_n$; it is given by
\begin{equation}
A_{mn} = H\!_{\lambda_m}\!(X_m\!\!-\!\!x_n, Y_m\!\!-\!\!y_n)\,K\!_{\lambda_m}\!(T_n)B\!_{\lambda_m}\!(T_n)\,\kappa(\lambda_m)\,\Delta\Omega_m\,.
\label{eq0}
\end{equation}
in which $H_\lambda(x,y)$ is the convolution of the point spread function (PSF) 
at wavelength $\lambda$ with the profile of an individual source component; 
$K_\lambda(T)$ represents a colour correction\footnote{In comparing the 
observed fluxes with model values, allowance must be
made for the fact the observations represent averages over finite
bandpasses rather than monochromatic values. Since the published PACS and 
SPIRE fluxes are based on an assumed source
spectrum which is flat in $\nu F_\nu$, allowance must therefore be made for 
other spectral shapes. For the present calculations, temperature-dependent
colour corrections are applied to the model images using the tables 
presented by \citet{pez13} and \citet{valt14}.};
$B_\lambda(T)$ is the Planck function; 
$\Delta\Omega_m$ is the solid angle subtended by the $m^{\rm th}$ 
pixel and $\kappa(\lambda)$ is the dust opacity 
law.  For present purposes we adopt a simple power law of the form
\begin{equation}
\kappa(\lambda) = 0.1\,{\rm cm}^2\,{\rm g}^{-1}\,\left(\frac{\lambda}{300\,\mu{\rm m}}\right)^{-\beta}\,.
\label{eq12}
\end{equation}
We use a power-law index, $\beta$, of 2, motivated by previous studies of
Galactic dust emission at {\it Herschel\/} wavelengths which find that SEDs
are fit well using this value (see \citep{sad12} and references therein).
The reference opacity (0.1 cm$^2$ g$^{-1}$ at 300 
$\mu$m) is defined with respect to total mass (dust plus gas). Although
observationally determined, it is consistent with a gas to dust ratio of 
100 \citep{hild83}. It should be noted that there is evidence for variations 
in dust properties throughout the Galaxy and such variations impact the
choices of both $\beta$ and the reference opacity. For example,
in the Galactic Centre region values of 1.2 and 0.042 cm$^2$ g$^{-1}$,
respectively, may be more appropriate \citep{rath15}. Dust properties have
also been found to differ between cool molecular clouds and warmer diffuse 
ISM \citep{cam01,para11}. While errors in the assumed reference opacity can
be rectified by simple scaling of the output column densities, this is
not true for the case of $\beta$ variations. We have, however, investigated
the effects of such variations and a quantitative example is presented in 
Section 4 below.

The {\it a priori\/} probability that a given cell in the state space is
occupied by a source component is controlled by
a ``dilution" parameter, $\eta$, defined as the ratio of the number of
source components to the total number of cells in the state space. The smaller 
the assumed value,
the more the algorithm tries to fit the data with the least possible amount of
source structure. The exact value is not critical, but
the most appropriate value is one which results in
a reduced chi squared value of order unity. 

The principal inputs to \ppmap\ are a set of observed images of 
dust continuum emission, their associated PSFs, the assumed dust opacity 
law over the wavelength range of the observations, and a grid of temperature 
values at which the differential column density will be estimated. 
The input images could include not only {\it Herschel\/} data but also,
where available, data from ground-based observatories, both single-dish
and interferometric. The output is an image cube of
differential column density of material (gas plus dust) per unit 
interval of dust temperature which, at each angular position, is expressed in 
units of hydrogen molecules per square centimetre per degree Kelvin.
Also included in the output is a corresponding image cube of uncertainty
values. Estimation errors are Poisson-like, and increase in regions of
high total column density. 

As discussed by \citet{mar15}, the advantages over
more conventional techniques for column density mapping are:
\begin{enumerate}
\item increased spatial resolution resulting from the incorporation of
PSF knowledge; all observational
images are used at their native resolution and it is not necessary
to smooth to a common resolution;
\item increased accuracy of peak column densities of compact features,
due both to the resolution improvement and to taking proper account of 
temperature variations along the line of sight;
\item the temperature decomposition provides the potential ability to
distinguish different physical phenomena superposed along the line of sight.
\end{enumerate}

\section[]{Observational data and analysis procedure}

We have used \ppmap\ to produce image cubes for all of the Hi-GAL data,
which consist of a set of {\em Herschel\/} PACS and SPIRE
images at wavelengths of 70, 160, 250, 350, and 
500 $\mu$m. 
The spatial resolution values of these data, i.e., the beam sizes at full 
width half maximum (FWHM), are approximately
8.5, 13.5, 18.2, 24.9, and 36.3 arcsec, respectively.
Details of the calibration and map-making procedures are
given by \citet{elia13}. Briefly, calibration for both instruments
was accomplished to Level 1
using routines in the {\it Herschel\/} Interactive Processing Environment
\citep[{\scriptsize HIPE};][]{ott10}, and subsequent map making utilised 
{\scriptsize UNIMAP} \citep{piazzo15}, which produced intensity maps
in units of MJy/sr for each band. For the SPIRE bands, the extended-source
flux calibration was used, based on band-averaged beam areas of 450, 795, 
and 1665 arcsec$^2$ at 250, 350, and 500 $\mu$m, respectively
\citep{griffin13}. Calibration of the absolute background levels
was achieved by applying a linear transform based on
coefficients determined by comparing {\it Herschel\/} with
{\it IRAS\/} and {\it Planck\/}, following \citet{bern10}.
The resulting uncertainty in absolute flux density is $\sim5$\% for all bands.

In running \ppmap\ on these data, we use PSFs
based on the measured {\it Herschel\/} beam profiles \citep{pog10,griffin13}. 
The adopted temperature grid consists of 12 temperatures equally spaced 
in $\log T_{\rm D}$ between 8 K and 50 K. It is designed to cover 
the expected range of temperatures to be encountered, based on an assessment of
the observed spectral energy distributions (SEDs). 
The actual number of temperature values is guided by the effective
temperature resolution, i.e., there should be a sufficient number of 
temperature values to ensure that the true temperature of a given source can
be approximated by at least one of the available temperatures. The only
penalty for using too many temperatures is that computational effort
increases proportionately. In this sense, the selection of the temperature
sampling interval is analagous to that of the spatial sampling interval. 
The 12 chosen temperatures represent discrete samples along the temperature 
axis---no averaging is done between samples. However, in order to
approximate the continuous function representing the
true temperature variation, the samples may be regarded
as the midpoints of a set of finite temperature intervals of which the 
$i^{\rm th}$ can be expressed as
\begin{equation}
\frac{T_{i\!-\!1}\!+\!T_i}{2} \,<\, T \,\le\, \frac{T_i\!+\!T_{i\!+\!1}}{2};
\label{eq01}
\end{equation}
for example, as implemented here, the third temperature bin, centred on
$T_3=13.2$ K, corresponds to the interval (12.2 K, 14.4 K).

The output spatial sampling interval for the 
present analysis is 6 arcsec per pixel, and
the spatial resolution of the output maps is $\sim12$ arcsec.
The latter represents the Nyquist limit for the quoted sampling interval and
corresponds to the minimum separation for which two adjacent 
point sources can be distinguished, even if the output image contains isolated 
peaks of smaller FWHM. It corresponds to a range of physical scales
$\sim0.06$--0.6 pc for the estimated range of distances, 1--10 kpc, involved in
the Hi-GAL survey.

Regarding the dilution parameter, $\eta$, we used a value 10.0 for the majority
of tiles. This value corresponds to a
situation in which multiple components may occupy a given cell.
The exact value, however, is not critical, and some improvement in resolution
was obtained in a few cases by reducing it to 0.3. Those cases involved 
the tiles centred at nominal longitudes, $\ell$, of $158^\circ$, $176^\circ$, 
$180^\circ$, $224^\circ$, $248^\circ$, and $316^\circ$. The effect, on the 
results, of the choice of $\eta$ and other parameters is discussed below, 
using the $\ell=17^\circ$ tile as an example.

\section[]{The VIALACTEA database: Illustrative results}

All 163 tiles of the Hi-GAL survey have been processed with \ppmap. Each tile
covers a $2.4^\circ\times2.4^\circ$ field and the complete set
spans the entire Galactic Plane.
The results, now publicly available via the website indicated in
footnote 1, consist of:
\begin{enumerate}
\item a total of 163 image cubes of differential column density with
6 arcsec spatial pixels and 12 values along the temperature axis, covering
the range 8--50 K in dust temperature;
\item corresponding image cubes of the uncertainties;
\item 2D maps of total column density and density-weighted mean dust
temperature, derived from the image cubes.
\end{enumerate}

\begin{figure*}
\includegraphics[width=150mm]{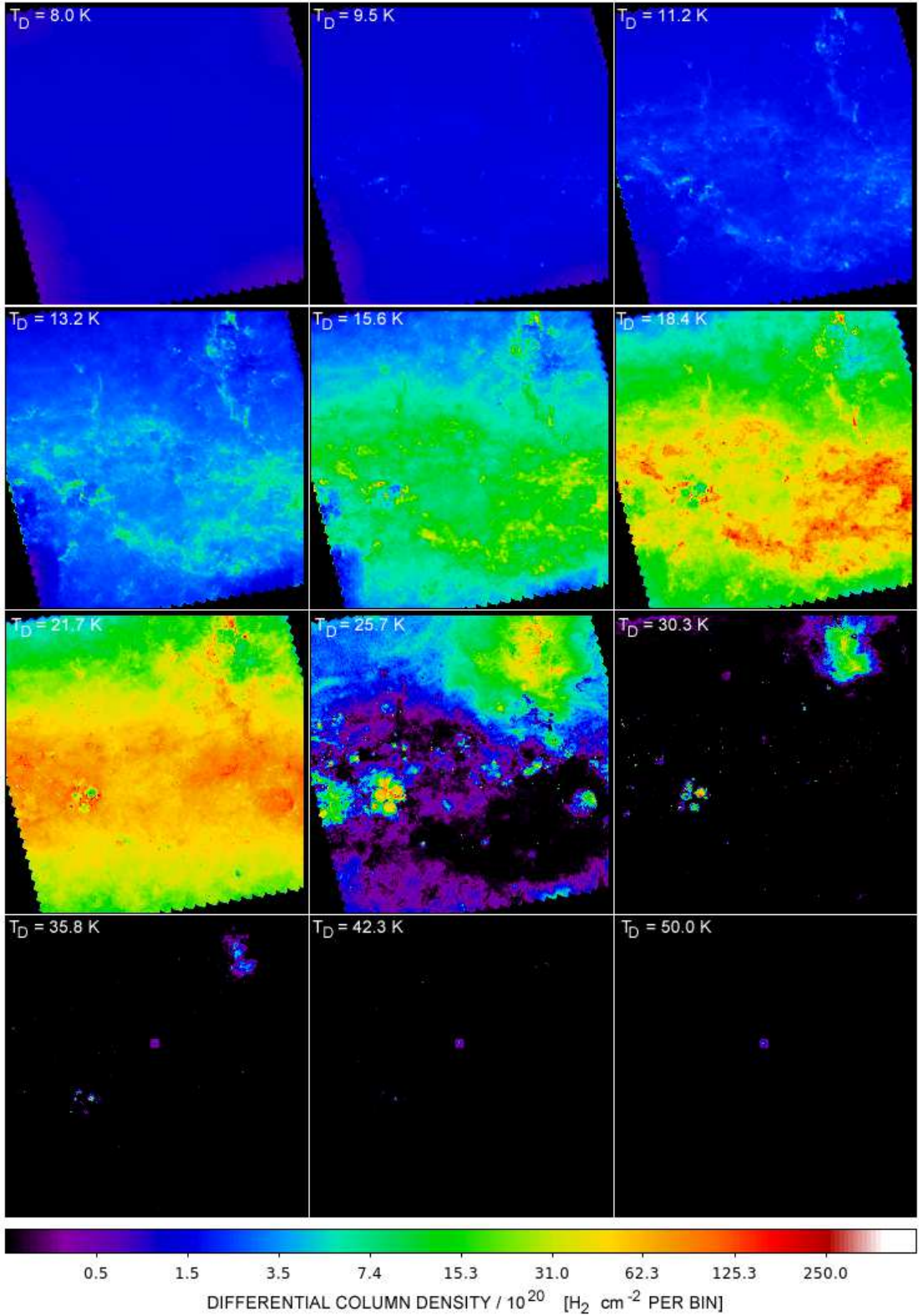}
\caption{Maps of differential column density, generated by \ppmap, for the
$2.4^\circ\times2.4^\circ$ field of the Hi-GAL tile centred on
Galactic longitude $\ell\simeq 17^\circ$. The field contains the Eagle Nebula 
(M16),
located in the upper right-hand quadrant. The corresponding dust temperature
is specified at the top left of each panel---it represents the midpoint 
of a finite temperature interval or ``bin" as defined in the text.}
\label{fig1}
\end{figure*}

\begin{figure*}
\includegraphics[width=150mm]{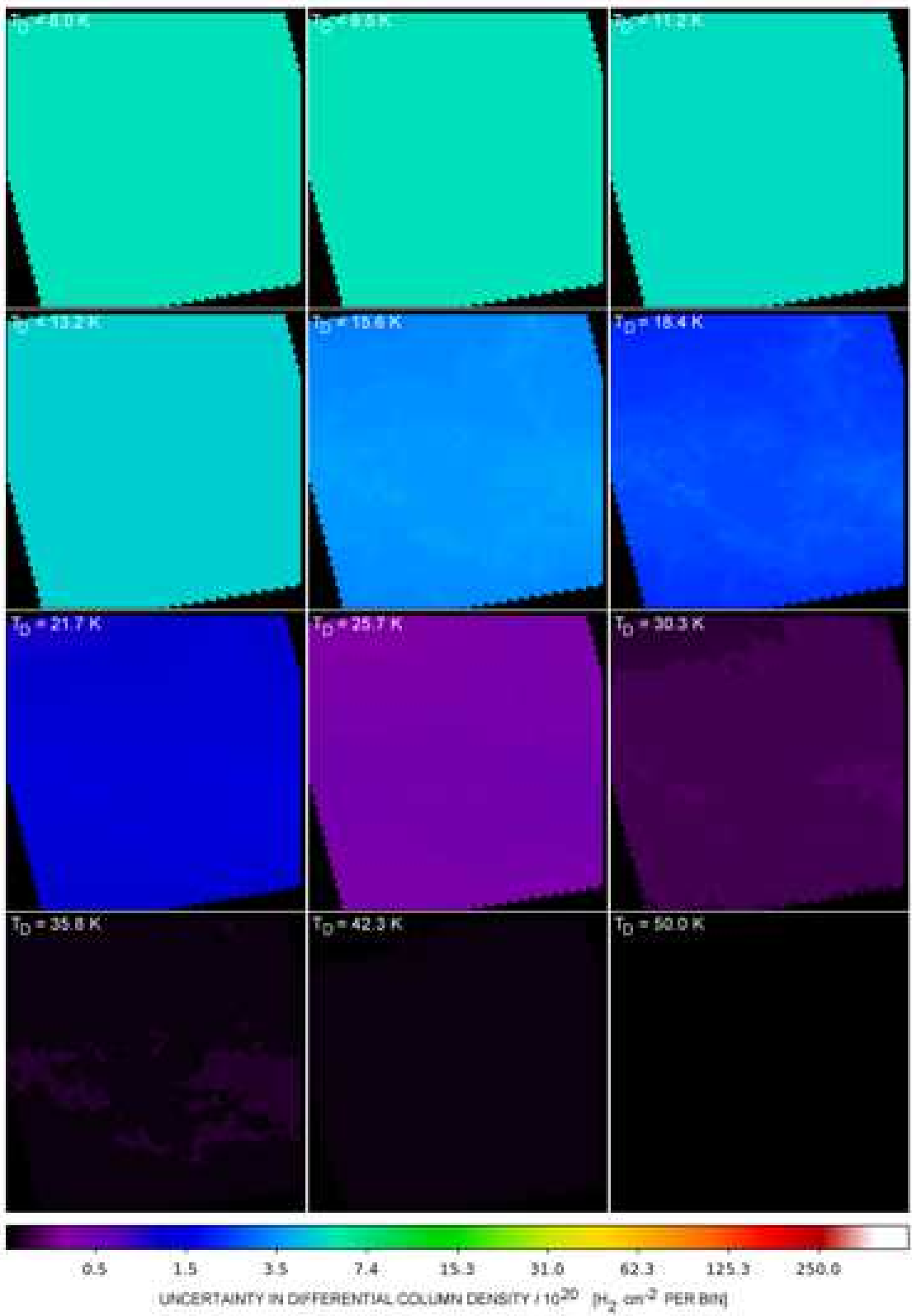}
\caption{Uncertainty maps associated with the results shown in Fig. \ref{fig1}.
Each represents the $1\sigma$ level of 
the corresponding map of differential column density.}
\label{fig2}
\end{figure*}

\begin{figure}
\includegraphics[width=90mm]{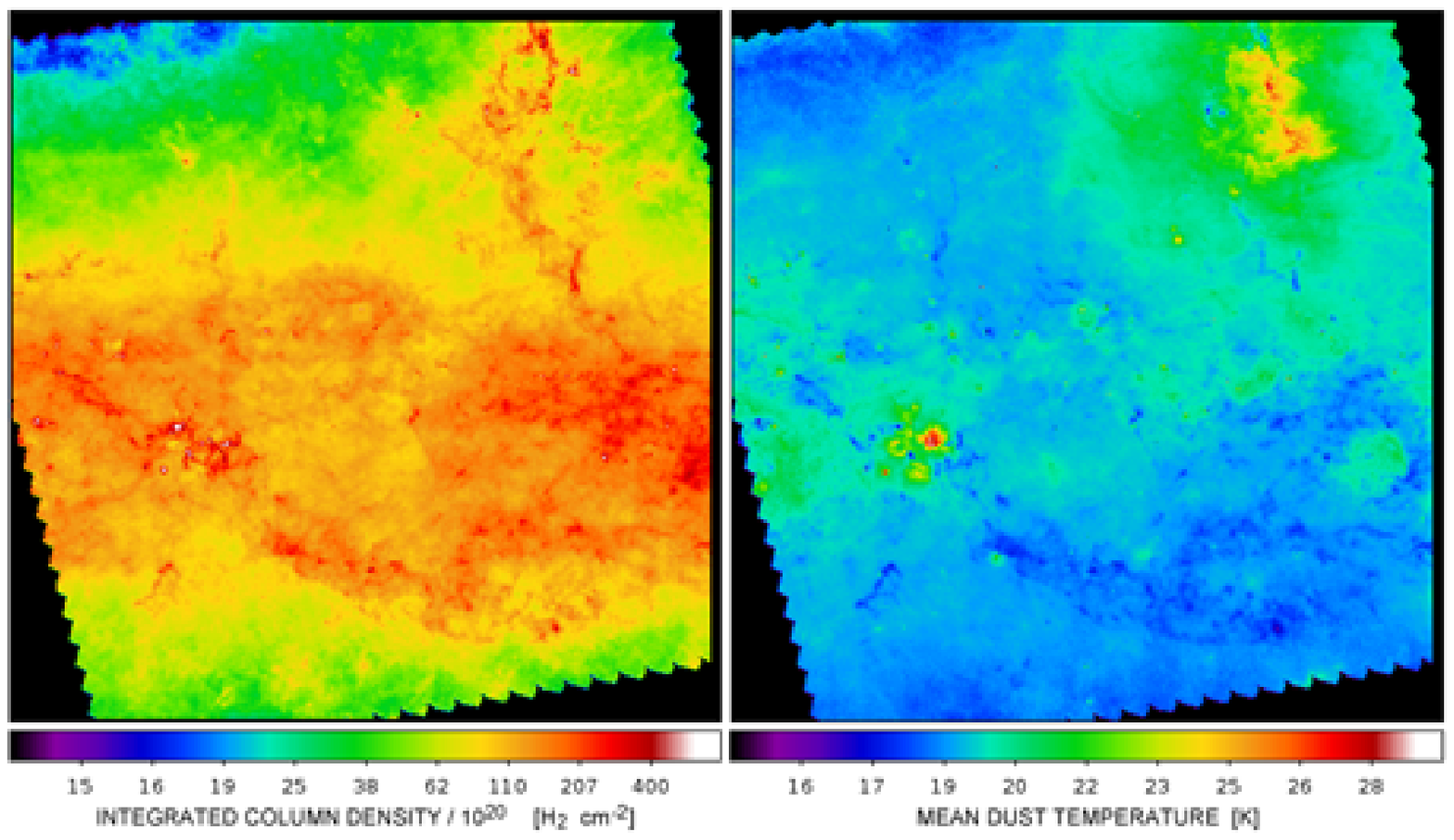}
\caption{Maps of integrated column density and mean dust temperature
for the $\ell\simeq 17^\circ$ Hi-GAL tile, derived from the \ppmap\ image
cube shown in Fig. \ref{fig1}.  The peak column density is $5.3 \times 10^{23}$
cm$^{-2}$, although the display scale is truncated at $8\times 10^{22}$ 
cm$^{-2}$ to improve the visibility of low-level features.}
\label{fig3}
\end{figure}

Figs. \ref{fig1}--\ref{fig3} illustrate the full set of results
for a representative tile centred at $\ell\simeq 17^\circ$. 
This field contains the Eagle Nebula (M16), a site of massive star formation
and the famous ``pillars of creation" \citep{hester96}. Its far infrared 
properties, as observed by {\it Herschel\/}, are discussed by \citet{hill12}.
Fig. \ref{fig1} shows
the differential column density in all 12 
temperature layers, and Fig. \ref{fig2} shows the associated uncertainties.
Clearly, the uncertainties are spatially quite uniform within each temperature 
bin, with no more than about a 50\% variation across the field. This indicates
that the errors are essentially background dominated,
and the sensitivity varies as a function of temperature based on the form
of the Planck function. The units of differential 
column density are ``hydrogen molecules per square centimetre per bin,"
where ``bin" refers to the temperature interval between adjacent midpoints as
defined in Eq. (\ref{eq01}). Fig. \ref{fig3} shows the maps
of integrated column density and mean (column density weighted) dust
temperature, both of which were derived from the image cube itself.

Uncertainties in differential column density are obtained using the
procedure specified in \citet{mar15} and are based on measurement errors
estimated from the sky background fluctuations in the individual 
observational images. They represent spatially uncorrelated noise and
do not include systematic effects such as flux calibration errors.
The correlation of those errors between bands could, in principle, result
in systematic effects in the variation of differential column density
between temperature bins. However, simulations by \citet{sad13} suggest
that such correlations affect the estimated dust temperatures by less
than 1 K. By far the largest source of uncertainty in the output maps arises
from the assumed value of the reference opacity which may be in
error by $\pm50$\%.

The output uncertainty maps themselves represent the $1\sigma$ level of 
the random component of differential column
density. Each pixel in these maps represents the {\it a posteriori\/}
standard deviation of the corresponding pixel in the differential column
density image. Examination of the map/uncertainty pairs shows that,
in some cases, the differential column density maps contain apparently
significant spatial structures which fall below the uncertainty level. 
This is due to
the finite resolution along the temperature axis, which allows structure
from one temperature plane to bleed through into an adjacent plane. As
discussed by \citet{mar15}, increasing the number of observational wavelengths
would increase the temperature resolution and hence reduce this tendency.

In order to investigate the sensitivity of \ppmap\ to variations in the
input parameters, we have considered three cases in addition to the
standard parameter set used to generate the results for the $\ell=17^\circ$
tile shown in Figs. \ref{fig1}--\ref{fig3}, namely:
\begin{enumerate}
\item changing the dust opacity index, $\beta$, from 2.0 to 1.5;
\item changing the number of temperature bins, $N_T$, from 12 to 8;
\item changing the dilution parameter, $\eta$, from 10.0 to 0.3.
\end{enumerate}
For each parameter set we have compared the results for the
$2.4^\circ\times2.4^\circ$ field based on five metrics,
namely the total cloud mass, peak column density, mean dust temperature,
and the minimum and maximum dust temperatures over the field.
The results are presented in Table 1, which
shows that the estimates are not strongly perturbed by the input parameter
variations. For example, the corresponding variations in estimated
cloud mass and peak column density are at the $\sim20$\% level, while
the temperature perturbations are within the
$\pm1$ K range corresponding to the typical temperature resolution
expected for {\it Herschel\/} data based on the simulations conducted by
\citet{mar15}.

\begin{table}
 \begin{minipage}{200mm}
  \caption{The effect of input parameter variations on the $\ell=17^\circ$ 
tile results.} 
  \begin{tabular}{@{}cccccccc@{}}
  \hline
$N_T$\footnote{Number of values in the temperature grid between 10 K and 50 K.}&
$\beta$\footnote{Opacity index.} & $\eta$\footnote{Dilution parameter.} &
$M_{\rm tot}$\footnote{Total mass of dust+gas in the $2.4^\circ\times2.4^\circ$ field.} &  Peak col. dens. & $\overline{T_{\rm D}}$\footnote{Density-weighted mean dust temperature along line-of-sight.} & $(T_{\rm D})_{\rm min}$ & $(T_{\rm D})_{\rm max}$\\
 &  & & $[M_\odot]$ & $[$H$_2$ cm$^{-2}]$ & $[$K$]$ & $[$K$]$ & $[$K$]$ \\
 \hline
  12 & 2.0 & 10.0 & $1.1\times10^6$ & $5.3\times10^{23}$ & 19.6 & 13.0 & 42.3 \\
   8 & 2.0 & 10.0 & $1.1\times10^6$ & $5.2\times10^{23}$ & 19.4 & 13.8 & 40.2 \\
  12 & 1.5 & 10.0 & $1.1\times10^6$ & $4.4\times10^{23}$ & 20.6 & 12.4 & 40.0 \\
  12 & 2.0 & 0.3  & $1.4\times10^6$ & $4.4\times10^{23}$ & 18.6 & 12.7 & 37.3 \\
\hline
\end{tabular}
\end{minipage}
\end{table}

We now discuss the results for two other regions in more
detail, specifically the W5-E bubble and the filamentary complex in
CMa OB1, and the radial variation of dust temperature in the Galaxy.

\subsection[]{W5-E}

\begin{figure}
\includegraphics[width=84mm]{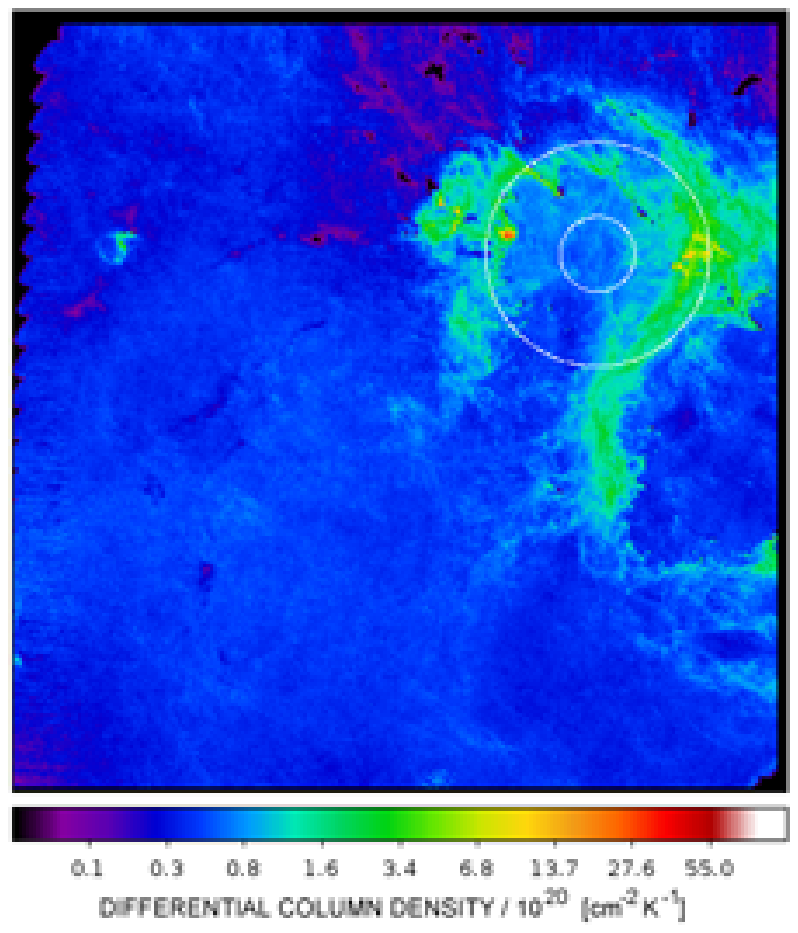}
\caption{Differential column density, at a dust temperature of 21.7 K,
over the $2.4^\circ\times2.4^\circ$ field of the Hi-GAL tile centred at 
$\ell\simeq 138^\circ$. The circles delineate subregions used in the calculation
of the column density histograms shown in Fig. \ref{fig10}.}
\label{fig4}
\end{figure}

The $\ell=138^\circ$ field is dominated by the Galactic HII region,
W5-E, which forms part of the W5 bubble complex
and is believed to be the site of triggered star formation
\citep{deharv2012}. It is ionised by a central O star and exhibits 
morphological features characteristic of 
HII regions which have expanded into surrounding neutral material. These 
include bright-rimmed clouds, BRC13 and BRC14, and inwardly-directed 
pillars, discussed by \citet{deharv2012}.
Fig. \ref{fig4} shows the differential column density image of the
full $2.4^\circ\times2.4^\circ$ tile in
the dust temperature slice centred at $T_{\rm D}=21.7$ K (i.e., 20.0 K to
23.6 K).

\begin{figure*}
\begin{center}
\includegraphics[width=180mm]{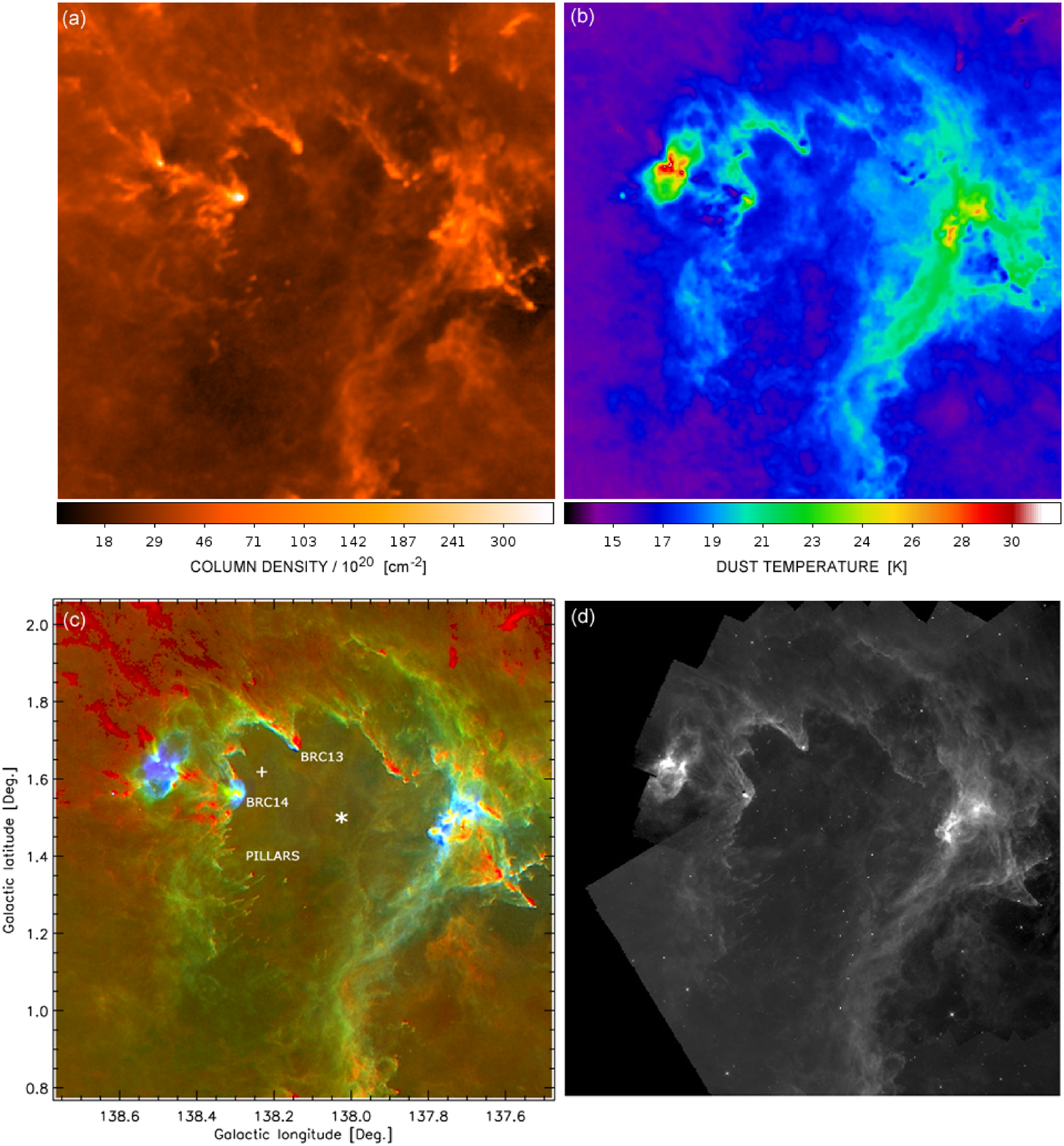}
\caption{The W5-E HII region bubble. Panels (a) and (b) show the results 
obtained by the conventional column density mapping procedure (see text).
Panel (c) shows the \ppmap\ results, represented as a 
composite map of differential column density in three different
temperature regimes, namely $T_{\rm D}\le 18.4$ K (red), 
$T_{\rm D}=21.7$ K (green), and $T_{\rm D}\ge 25.6$ K (blue). The ionising
star HD 18326 (O7V) is indicated by the asterisk.
Another hot star in the field is V1018 Cas (O7V--B1V), indicated
by the ``+" symbol. For comparison, (d) shows the 8 $\mu$m image from
{\it Spitzer\/}. Note how closely it resembles the distribution of 21.7 K dust 
shown in (c).}
\label{fig5}
\end{center}
\end{figure*}

Mapping of this field using the
conventional\footnote{For this example we have adopted the
procedure and detailed parameters specified by \citet{kon10}, using images at
all five {\it Herschel\/} continuum wavelengths, smoothed to 
36 arcsec resolution, a pixel size of 6 arcsec, and the dust opacity law given 
by Eq. (\ref{eq1}).} approach provides the images in panels (a) and (b)
of Fig. \ref{fig5}, representing integrated column density and mean
dust temperature. In that technique, the dust temperature is assumed to 
be constant along the line of sight and the data are first smoothed to a 
common spatial resolution. 
For comparison, panel (c) of Fig. \ref{fig5} shows the results of 
applying the \ppmap\ technique to the
same data. It is a composite image of W5-E showing the distribution of
differential column density in three different dust temperature regimes,
namely $T_{\rm D}\le 18.4$ K, $T_{\rm D}=21.7$ K, and $T_{\rm D}\ge 25.6$ K. 
Overplotted on this figure are the locations of the bright-rimmed clouds, 
BRC13 and BRC14, and the two hot stars, HD 18326 and V1018 Cas. 

\begin{figure}
\begin{center}
\includegraphics[width=84mm]{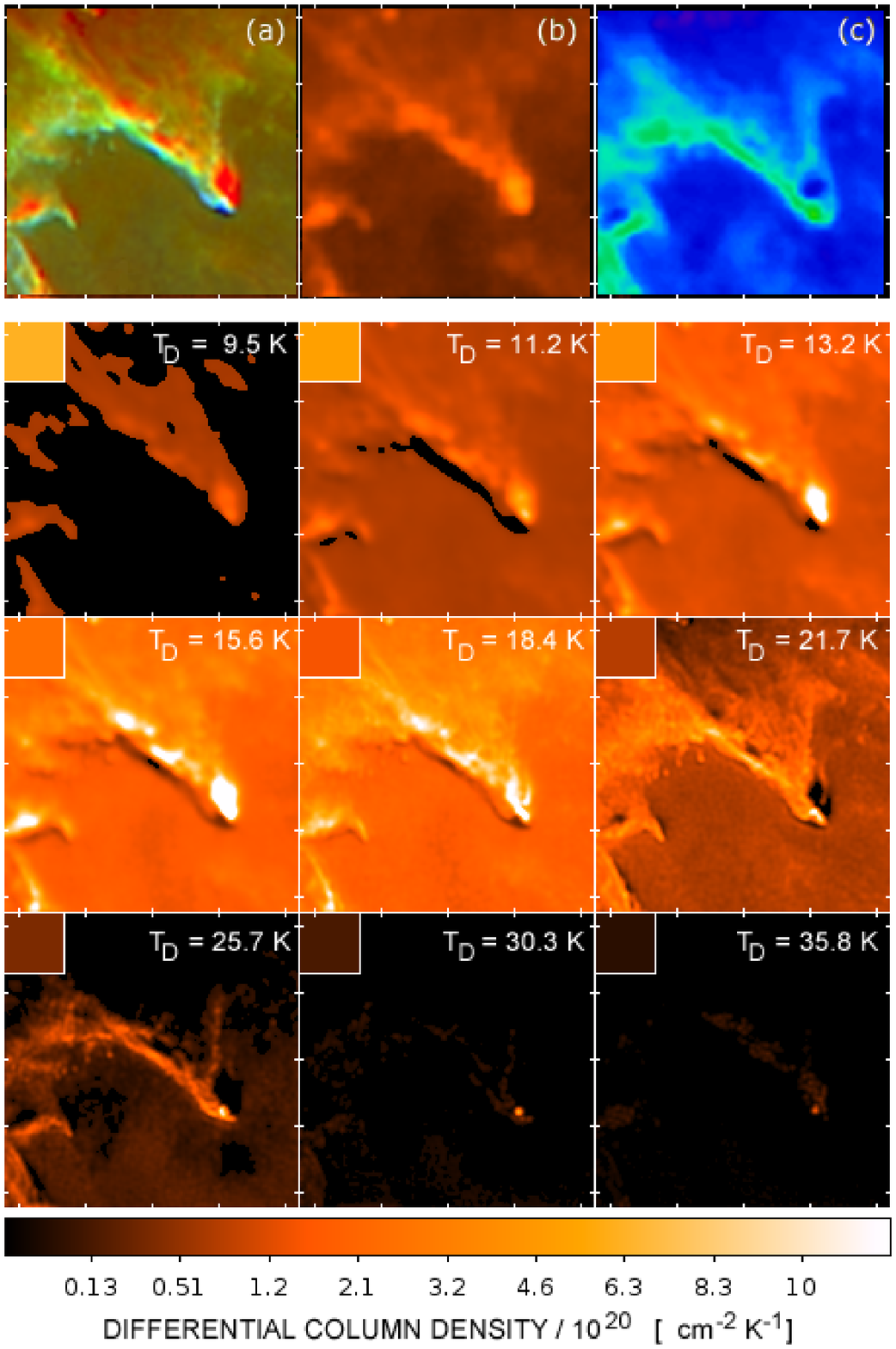}
\caption{Zoom-ins of the BRC13 column in W5-E.
{\it Upper row:\/} (a) Three-temperature composite produced by \ppmap, cut out 
from the bottom panel of Fig. \ref{fig5};
panels (b) \& (c) show the results
of the conventional mapping procedure in the form of integrated column
density and mean dust temperature, respectively.
{\it Lower 3\/} $\!\!\times\!\!$ {\it 3 block:\/}
Differential column-density, estimated by \ppmap,
in nine different planes of dust temperature, $T_{\rm D}$,
as indicated in the top right of each panel. In each case, the box in the top 
left indicates the $1\sigma$ uncertainty level. All panels are
presented on the same scale of differential column density.  The field of view
in each case is $15'\times15'$, centred on $(\ell,\,b)=(138.2052^\circ,\,
1.7260^\circ)$, and the pixel size is 6 arcsec
(0.058 pc at the assumed 2 kpc distance). 
The tick marks are at intervals of 2 pc.}
\label{fig6}
\end{center}
\end{figure}

Fig. \ref{fig6} (lower nine frames) shows the \ppmap-derived differential 
column density in nine separate temperature planes spanning the range
9.5--35.8 K, for a small field of view ($15'\times15'$)
around BRC13. It reveals that at the lowest temperatures ($T_{\rm D}
\stackrel{<}{_\sim}11$ K) the only visible structure is
a compact feature presumably representing a cool condensation at the tip
of BRC13. Proceeding to higher temperatures, the column-like morphology of
BRC13 becomes more apparent. For
$T_{\rm D}\sim11$--18 K a dark feature is visible along the eastern
edge representing a deficit of material in that temperature range, but this edge
becomes bright at $T_{\rm D}=25.7$ indicating that hotter material has
replaced cool material there.
Moving to still higher temperatures we find that at $T_{\rm D}=30.3$ K and
35.8 K, the dust distribution is dominated by a warm compact structure at the
southwestern tip of the column, suggesting a protostar which
has formed in the condensation. It does, in fact, correspond with a
young stellar object (YSO), of spectral class B5, whose ID is 16614 in the 
W5 IR Excess Spectral Catalog of \citet{koenig11}.
For comparison, the upper portion of Fig. \ref{fig6} shows the corresponding
three-temperature composite extracted from Fig. \ref{fig5}, and also
a pair of lower resolution maps (integrated column density and mean temperature)
produced by the conventional mapping technique.

\smallskip
Some interesting features revealed by Figs. \ref{fig5} and \ref{fig6}  are:
\begin{enumerate}
  \item the strong temperature gradient along the eastern edge of BRC13
which suggests heating from a direction other than the central O star.
One possible heating source is the YSO
at the tip of BRC14. Another is the hot star V1018 Cas. Although the latter
is not regarded as a member of the W5 complex due to its $\sim40$--50 km
${\rm s}^{-1}$ offset in radial velocity, it is conceivable that this
offset could arise if the bubble complex were the result of a cloud-cloud
collision via the model of \citet{habe92} in a similar fashion as for the
RCW 120 bubble \citep{tor15}. Detailed radiative transfer modelling of
BRC13, with
constraints provided by the source luminosities and the estimated temperature 
gradient, may serve to identify the heating source and therefore potentially
provide clues to the origin of the bubble complex;
  \item the radially-directed ``pillars" discussed by \citet{deharv2012},
each of which has a cool condensation at its tip;
  \item the morphology of the 22 K material (shown in green in 
panel (c) of Fig. \ref{fig5})
which has strong spatial correspondence with 8 $\mu$m emission imaged
by {\it Spitzer\/}, shown in panel (d) of the same figure. That emission 
is believed to
arise from PAH particles in the surrounding photodissociation region (PDR)
which fluoresce in the ultraviolet radiation from the O star 
\citep{deharv2012}.
This correspondence suggests that the PDR is characterised by dust
at $\sim22$ K, enabling its spatial structure to be distinguished
from that of other dust emission components along the line of sight.
\end{enumerate}

\begin{figure}
\includegraphics[width=84mm]{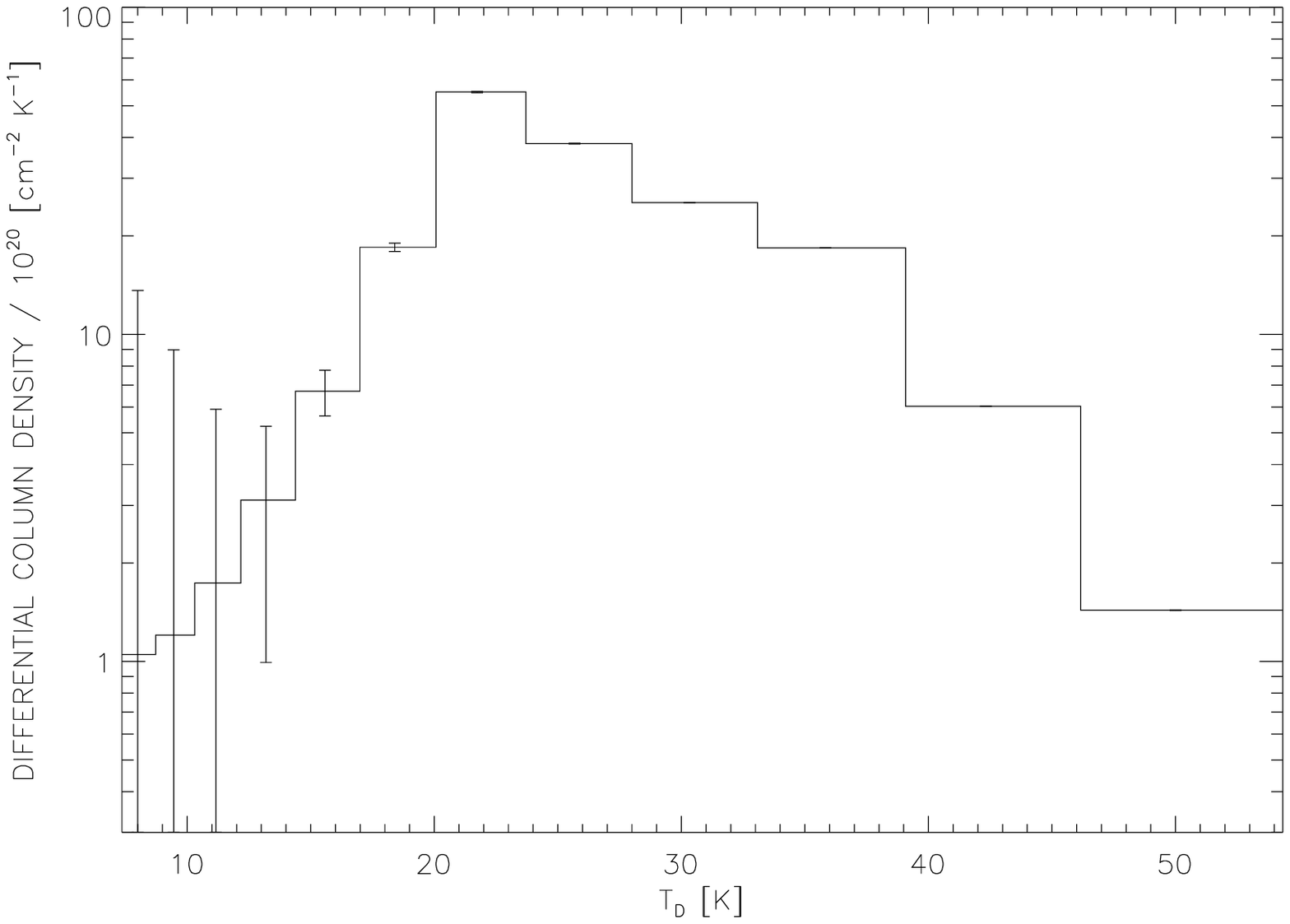}
\caption{Profile of differential column density in W5-E as a function of dust 
temperature, at the location of the YSO at the tip of BRC14.}
\label{fig7}
\end{figure}

The multiple-temperature planes of the \ppmap\ image cube enable the
temperature profile to be extracted at each spatial pixel location.
An example is presented in Fig. \ref{fig7} which shows the profile of
differential column density as a function of temperature at the location
of the YSO at the tip of BRC14. It shows that, on this line of sight,  
significant amounts of dust exist at all temperatures between 15 K and 50 K.
Hence the adoption of a single (mean) temperature is a very poor approximation.

\begin{figure}
\center
\hspace*{-0.5cm}\includegraphics[width=81mm]{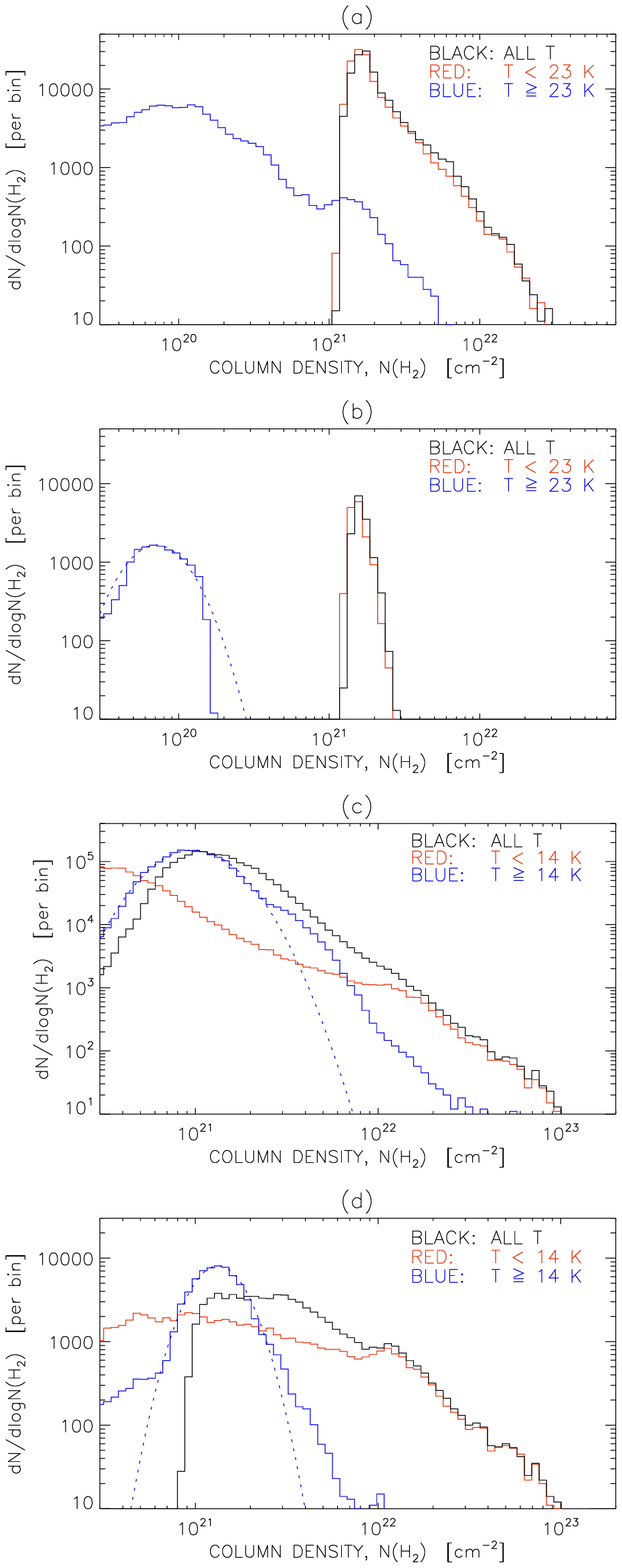}
\caption{Column density histograms for 
W5-E and CMa OB1, as follows:
(a) Overall bubble structure of W5-E; (b) Central part of W5-E. Details
of the actual subregion boundaries are given in the text.
Total column density is shown in black, while the coloured curves represent
temperature ranges as indicated. 
(c) Full $2.4^\circ\times2.4^\circ$ field 
containing CMa OB1; (d) Central filamentary complex in CMa OB1.
The dotted lines represent least squares
fits of lognormal functions to the histograms of the warm components.
Note that, although the total of the black histogram values must equal
the combined total for the red and blue histograms in each case, there is no
requirement that the black histogram should equal or exceed the
red or blue histogram at individual values of column density, as discussed
further in the text.}
\label{fig8}
\end{figure}

Fig. \ref{fig8} shows histograms of column density for
portions of W5-E. Panel 8(a) is derived from
pixels within a circular region of diameter 24 pc centred on the exciting star,
corresponding to the bubble boundary as assessed by
\citet{deharv2012}. It is represented by the larger of the two circles
overplotted in Fig. \ref{fig4}. Panel 8(b) is based on a smaller region
corresponding to the smaller circle in Fig. \ref{fig4}, representing
lines of sight through the centre of the HII region rather than the 
peripheral structures. 
The reason that the blue histograms exceed the black (total) 
histograms at the low-density end is that, while the total column density 
includes contributions from warm low-density material, all lines of sight
intersect higher density material somewhere along the way and thus the
low density bins of the ``total" histogram are unpopulated. 

The histograms in (b) have a fairly simple interpretation
in terms of a warm component whose column densities are distributed
approximately as a lognormal, plus a cool component with a narrow range
of column densities. The narrowness of this ``spike" is probably the result
of looking along the normal through the thin shell comprising the PDR.
This indicates that some features in column density histograms can be the
result of systematic effects related to geometry rather than representing
purely the probability distribution of random density fluctuations related
to turbulence. Interestingly, the spike occurs at the same column density 
as the peak of the broader distribution of cool dust in panel (a).
This suggests the possibility that the broad apparent power-law tail
might be decomposed into separate narrow components of which the spike in (b)
represents the low column density limit. Further analysis may thus shed light
on whether the PDFs of star-forming clouds can be represented better by
turbulence-induced
lognormal distributions in the presence of noise \citep{brunt15},
superpositions of lognormals from different parts of the field,
combinations of lognormals and gravity-induced
power-laws \citep{froe10,sch12, sch13}, or truncated power laws \citep{lom15}.
The distinction is important since it bears on the question of whether
probability distribution functions (PDFs) of column density
provide a probe of self-gravity effects in star-forming bubbles such
as W5-E.

\subsection[]{CMa OB1}

\begin{figure}
\includegraphics[width=84mm]{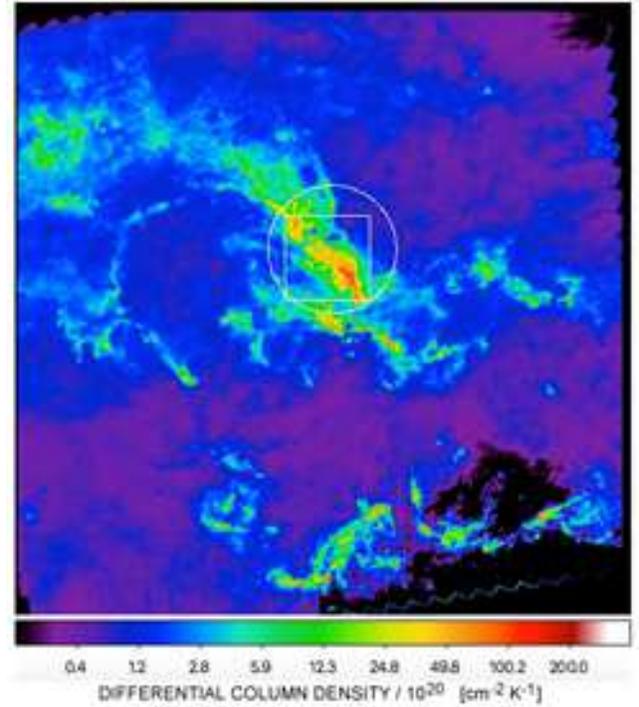}
\caption{Differential column density, at a dust temperature of 13.2 K,
over the $2.4^\circ\times2.4^\circ$ field of the Hi-GAL tile at 
$\ell\simeq 224^\circ$. The circle delineates a subregion used in the 
calculation of column density histograms shown in Fig. \ref{fig8}.
The square represents the boundary of the cutout region shown in Fig. 
\ref{fig10}.}
\label{fig9}
\end{figure}

\begin{figure*}
\includegraphics[width=120mm]{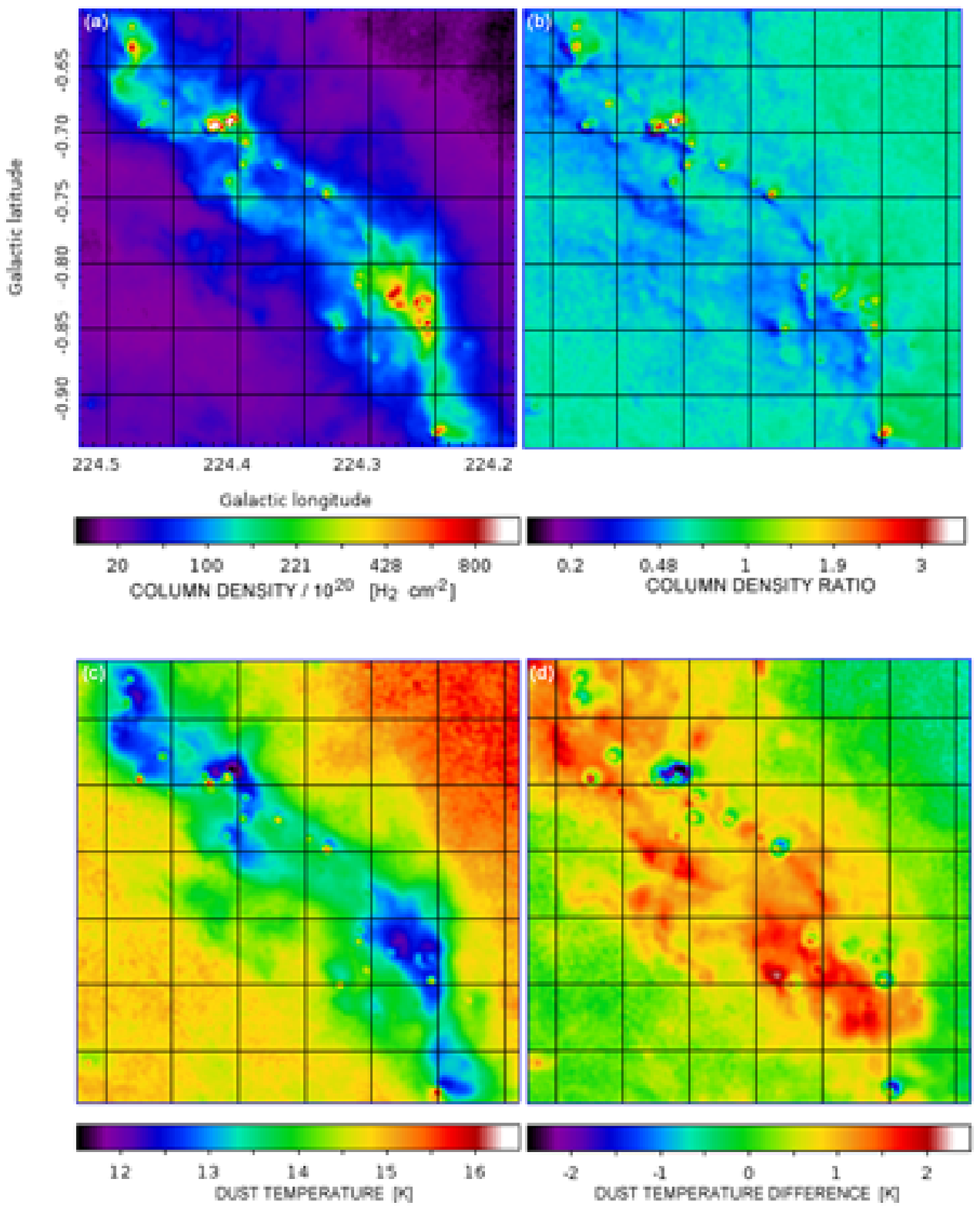}
\caption{Comparison between results of \ppmap\ and the ``conventional"
column density mapping
technique (see text) for a $20'\times20'$ subfield of the 
$\ell\simeq 224^\circ$ Hi-GAL tile (delineated by the square region in
Fig. \ref{fig9}), showing part of the CMa OB1 filamentary 
complex.  (a) Integrated column density
from \ppmap; (b) column density ratio (\ppmap\ divided by ``conventional");
(c) mean dust temperature; (d) dust temperature difference (\ppmap\ 
minus ``conventional").}
\label{fig10}
\end{figure*}

The $\ell=224^\circ$ tile is dominated by the large filamentary complex of
CMa OB1, a site of prolific star formation \citep{elia13}. The spatial
distribution of temperature-differential column density at $T_{\rm D}=13.2$ K,
produced by \ppmap, is shown in Fig. \ref{fig9}. Maps of integrated column
density and density-weighted mean dust temperature for a $20'\times 20'$
subregion are shown in Fig. \ref{fig10} (panels a and c, respectively).
Panels 10(b) and 10(d) enable
a comparison with the results of the ``conventional" technique discussed 
earlier. Specifically panel 10(b) shows the ratio of
estimated column densities (\ppmap\ divided by ``conventional"), while
panel 10(d) represents the temperature {\it difference\/} (\ppmap\ minus
``conventional"). While the two techniques give comparable results for
large areas of the background, \ppmap\ gives larger estimates of column
density at the peaks of compact sources (by factors of up to 6). It also
gives lower estimated temperatures in cool compact sources (cores) and
larger peak values for the warm compact sources (protostars). These differences
result partly from the increased spatial resolution provided by 
deconvolution, and also from the increased accuracy
obtained by dispensing with the uniform temperature assumption. Interestingly,
the \ppmap\ estimate of total mass within the $20'\times20'$ region 
(4700 $M_\odot$) is significantly smaller (by a factor of 0.6) than the 
value 7700 $M_\odot$ yielded by the ``conventional" method. 
Much of this discrepancy can, however, be attributed to differences in dust 
temperature (13.7 K and 13.0 K) as estimated by the two techniques, 
respectively.

The full image cube for the main part of the filamentary complex has been 
presented by \citet{mar15}.  It enabled the PDF
of column density to be decomposed by 
temperature, showing the warm ($T_{\rm D}>13$ K) and
cool ($T_{\rm D}<13$ K) material to be characterised by different
distributions of column density. Specifically, the PDF of the warm
material was found to be well approximated by a lognormal, as expected
for density fluctuations resulting from interstellar turbulence, while 
that of the cool material was consistent with power law behaviour 
due to the effects of self-gravity on those fluctuations 
\citep{klessen01,kain09,krit11,sch13,giri14}.

The lower two panels of Fig. \ref{fig8} show column density histograms,
representing estimates of PDFs, derived
for the $\ell=224^\circ$ tile using \ppmap. Panel (c) is for the full
$2.4^\circ\times2.4^\circ$ field, while panel (d) is for a more limited
region enclosing the main part of the filamentary complex as indicated
by the circular boundary in Fig. \ref{fig8}. The histograms of total
column density (shown in black) have been decomposed into contributions from 
warm ($T_{\rm D}\ge14$ K) and cool ($T_{\rm D}<14$ K) material, represented
by the blue and red histograms, respectively. 
Clearly, the warm material exhibits
lognormal-like behaviour while the cool material accounts for the
high density tail. The separate histograms for the smaller 
subregion illustrate the effect of field selection on the
shape of the histogram. The circular boundary in this case is intended to
delineate the region which appears to contain the majority of star
formation activity, based largely on visual inspection. An interesting line
of research to be pursued in the future is to explore how the PDFs of
temperature-differential column density vary between different subregions
and between concentric regions of different size,
as has been done previously with PDFs of {\it total\/} column density 
\citep{rus13,trem14,sch15a,sch15b}.

\subsection[]{Radial profile of dust temperature in the Galaxy}

Estimates of dust temperature can provide strong constraints on the
interstellar radiation field (ISRF), which is of interest not only for
star formation models, but also for the propagation of cosmic rays in
the Galaxy \citep{por05}. In a previous study, \citet{sod97} used data from
the DIRBE instrument on {\it COBE\/},
in combination with $^{12}$CO, HI, and radio continuum data, to
estimate the dust temperature as a function of Galactocentric distance
for the dust embedded in neutral atomic (HI), molecular (H$_2$), and
ionised (HII) hydrogen. In all cases they found that the dust temperature
decreases with radial distance, but with large error bars and relatively
low spatial resolution. 

Here we report the first {\it Herschel\/}-based estimate of the radial profile 
of dust temperature in the Galactic Plane. This is accomplished by
first calculating the mean 
dust temperature (weighted by differential column density)
using the image cube generated by \ppmap\ for each 
$2^\circ\times2^\circ$ Hi-GAL tile. Then, in order to assign a characteristic
heliocentric distance to each tile, a Galactic spiral arm model is
needed. We base our heliocentric distances on the results presented by 
\citet{reid16}, which make use of trigonometric parallaxes from VLBI
observations of water and methanol masers in high mass star formation
regions, supplemented with other data such as kinematic information from
CO observations. Where necessary, ambiguities in spiral arm identification
are resolved with the aid of the peaks in source counts associated with
tangential directions, in a fashion similar to that discussed by \citet{rag16}.
The heliocentric distances are then converted to Galactocentric distances
assuming a distance to the Galactic centre of 8.4 kpc \citep{reid09}.

\begin{figure}
\includegraphics[width=84mm]{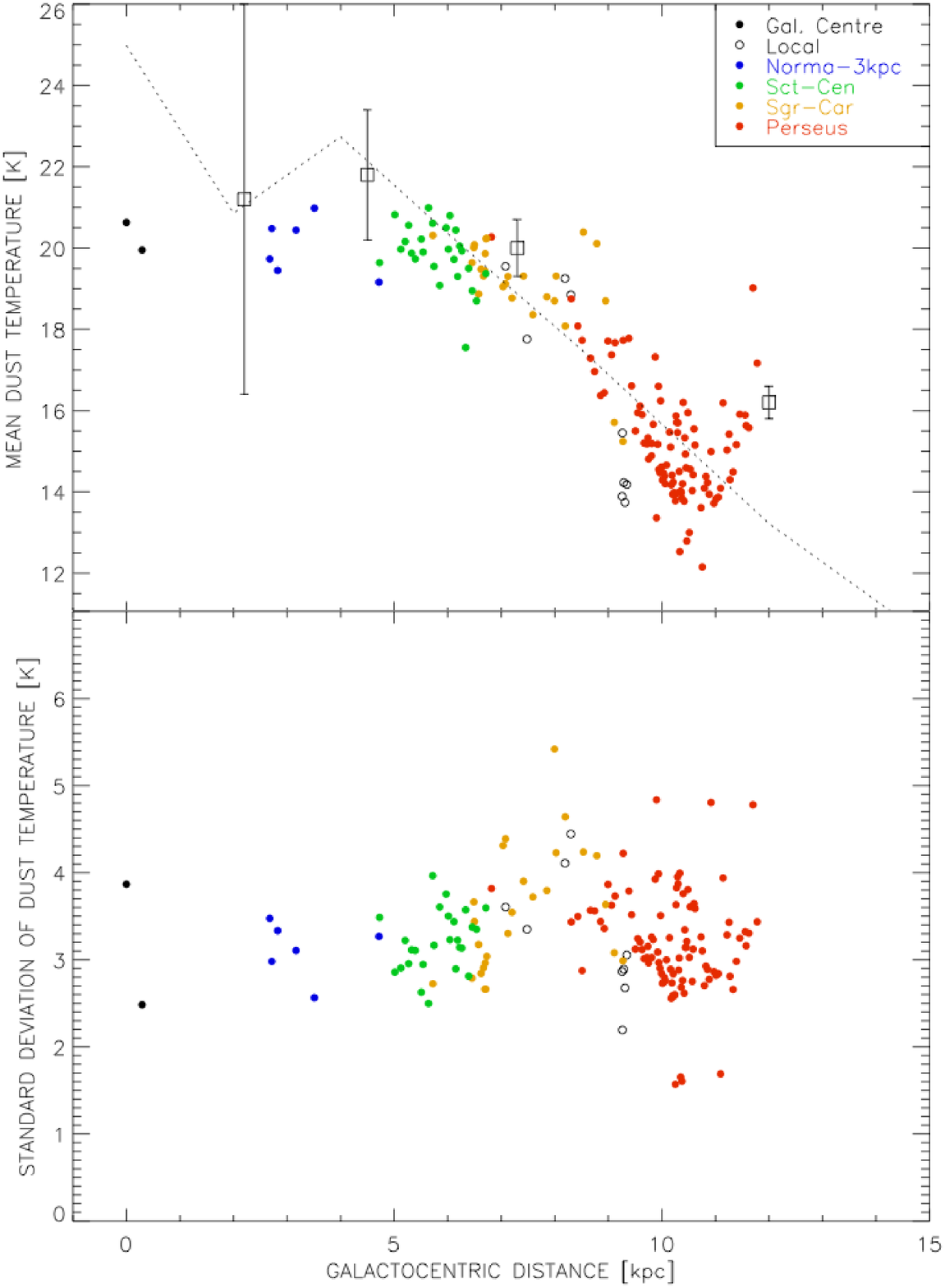}
\caption{{\it Upper panel:\/} mean dust temperature as a function of 
Galactocentric distance.
The colour coding represents the assumed spiral arm indentifications based on
the results presented by \citet{reid16}, as indicated in the upper right of
the panel.
The dotted line represents the model curve based on the ISRF of
\citet{por08}. The open squares with error bars are from \citet{sod97}.
{\it Lower panel:\/} the standard deviation of dust temperature values
(along the line of sight as well as in the plane of the sky) as a function of 
Galactocentric distance.}
\label{fig11}
\end{figure}

\begin{figure}
\includegraphics[width=84mm]{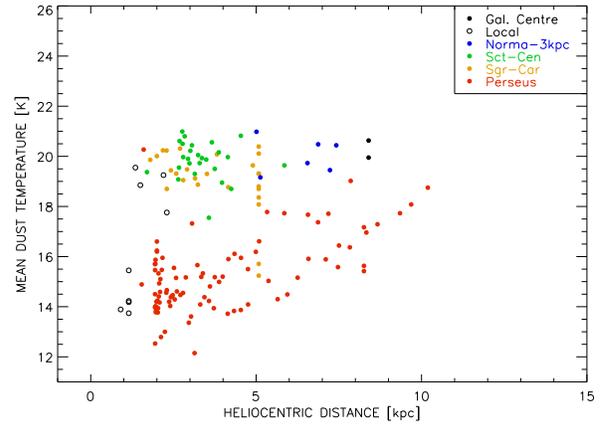}
\caption{Mean dust temperature as a function of heliocentric distance,
to be compared with the variation with respect to {\it Galacto\/}centric
distance shown in Fig. \ref{fig11}.}
\label{fig12}
\end{figure}

The top panel of Fig. \ref{fig11} shows a plot of mean dust temperature as a 
function of
Galactocentric distance. It exhibits a clear trend of decreasing dust
temperature with increasing distance. To check whether this trend could be 
an artifact due to differing {\it heliocentric\/} distances,
Fig. \ref{fig12} shows the corresponding plot as a function of heliocentric
distance instead. The lack of correlation indicates that the estimated dust
temperatures depend on Galactocentric rather than heliocentric distance.
We note also that our {\it Herschel\/}-based estimates are reasonably consistent
with those of \citet{sod97} which are shown overplotted on Fig. \ref{fig11}.

The image cubes serve to characterise not just the mean dust temperature,
$\overline{T_{\rm D}}$, but
also higher order moments of the dust temperature along the line of sight. The
lower panel of Fig.
\ref{fig11} shows the standard deviation of dust temperature values within
the image cube for each Galactocentric distance. The median value is 3.2 K,
and there is no significant trend with Galactocentric distance.
Similarly, we find no pronounced
trends with Galactocentric distance for the normalised skewness and kurtosis, 
for which the median values are $-0.3$ and 3.5, respectively. Those
quantities are defined as the density-weighted mean values of
$(T_{\rm D}-\overline{T_{\rm D}})^3$ and $(T_{\rm D}-\overline{T_{\rm D}})^4$, 
divided by the variance to the power
of 3/2 and 2, respectively; 
their theoretical values for a pure Gaussian distribution are 0 and 3.0, 
respectively. The \ppmap\ results thus serve to characterise the overall
temperature variation in the Galactic Plane as essentially a Gaussian with 
constant variance such that the only significant variation with Galactocentric
distance is in the mean value.

It is of interest to ask whether the observed variation of mean dust temperature
with radial distance, $r$, is consistent with heating by the ISRF. We can
compare our estimates in Fig. \ref{fig11} with a model prediction by noting
that, based on the ISRF heating model, the dust temperature $T_{\rm D}(r)$
should vary approximately as 
$I_{\rm ISRF}(r)^\frac{1}{4+\beta}$, where $\beta$ is the dust
opacity index \citep{bern10} and $I_{\rm ISRF}(r)$ is the bolometric
intensity of the ISRF. Assuming $\beta=2$, using the 
wavelength-integrated ISRF intensities from \citet{por08}\footnote{
Downloaded from http://galprop.stanford.edu/} and the
{\it COBE\/} measurement of mean dust temperature in the solar neighbourhood,
17.5 K \citep{lag98}, we obtain the model curve represented by the dotted
line in Fig. \ref{fig11}. The model curve is largely consistent with
the distribution obtained from the Hi-GAL data, and with
the previous observational estimates from \citet{sod97}, but there are
some qualitative differences as discussed below. 

\section[]{Discussion}

These results serve to illustrate the information which can be gained
from the line-of-sight temperature decomposition provided by \ppmap\ in
the study of dusty Galactic structures on all scales. The main assumption
involved is that the dust emission is optically thin. This is valid
over most of the Galactic Plane at the {\it Herschel\/} wavelengths---the
fraction of output pixels for which the inferred optical depth exceeds unity
is $\sim10^{-4}$ at 70 $\mu$m and $\sim10^{-6}$ at 160 $\mu$m. Thus violations
of the optically thin assumption have negligible effect on estimates of
the large-scale variation of dust temperature in the Galaxy. There are,
however, localised regions of high 70 $\mu$m optical depth in dense 
molecular clouds, and one might expect an impact on estimates of 
peak column density and core masses. However, a study of the dense molecular
cloud `The Brick' by \citet{mar16} suggests that high optical depth at
a single wavelength (70 $\mu$m) does not significantly perturb the solution,
provided there is sufficient signal to noise at the longer wavelengths.
Nevertheless, we cannot discount the possibility that there exist some very 
dense cores for which optical depth effects would result in underestimates 
of column density and hence core mass.

For the star-forming HII region, W5-E, the temperature
decomposition facilitates the separation of distinct physical components 
along the line of sight. One example is the image of 22 K 
dust which apparently represents the morphology of
the PDR surrounding the HII region, whereas images at lower temperatures
are dominated by cooler, clumpy surrounding material; 
those at higher temperatures are dominated by features which
represent protostars and possibly a bow shock.
All of these phenomena co-exist along some of the
same lines of sight but are separable in different temperature planes.
In a similar fashion, the temperature-dependent images of differential 
column density in CMa OB1 provide separate views of
the cool filamentary material and the warmer overlying material of the ISM.
The W5-E results also reveal the existence
of a strong temperature gradient along one edge of a column-like protrusion
which requires a heating source other than the ionising star. If subsequent
radiative transfer modelling reveals that source
to be the O7/B1 star V1018 Cas, it would strongly suggest the latter to be
a member of the W5-E complex despite its large offset in radial velocity.
This might be due to W5-E having been formed in a cloud-cloud collision
\citep[e.g.,][]{habe92} or it might be the result of a dynamical sling-shot
ejection. The ubiquitousness of such bubbles in the Galactic Plane 
\citep[see, for example,][]{and11} means that the set of \ppmap\ image
cubes from Hi-GAL should provide a large enough statistical sample to
draw some more general conclusions regarding the possible role
of phenomena such as cloud-cloud collisions in the formation of these
objects \citep[e.g.,][]{bal15,bal17}.

Although our estimated radial variation of dust temperature in the 
Galaxy can be explained broadly in terms of heating of the dust grains by the 
ISRF (see dotted line on Fig. \ref{fig11}), 
not all features are reproduced by this simple heating model.
The estimated variation appears to be characterised by two regimes, 
one in which the temperature is approximately constant out to $\sim6$ kpc,
and the other which exhibits a relatively steep falloff beyond that. 
The transition point appears to coincide with
the sudden increase in the proportion of molecular with respect to neutral
hydrogen in the inner Galaxy \citep{sod97}, and might therefore be understood 
in terms of the shielding effects of dust in dense
molecular clouds. 

The \citet{sod97} results suggest that for most of the range in
radial distance, the dust component dominating our temperature estimates is 
located in the diffuse HI clouds of the Galaxy. In a forthcoming paper
we will use information gathered from compact source extractions to help
distinguish between the Galactic temperature profiles of the diffuse 
cloud component and the denser star-forming clouds. We will also investigate 
possible connections with radial trends
in star formation rate \citep[see, for example,][]{rag16}.

\section[]{Conclusions}

The \ppmap\ technique provides image cubes of differential column density
as a function of dust temperature and ($x,y$) sky location and represents
a clear advance in our ability to extract information on dust distributions 
from continuum observations at multiple wavelengths. It thus represents
a very powerful tool for the analysis of long-wavelength dust emission
data from {\it Herschel\/} and other far-infrared and submillimetre
facilities. Specific advantages are:
\begin{enumerate}
\item increased spatial resolution resulting from the incorporation of
PSF knowledge; all observational
images are used at their native resolution and it is not necessary
to smooth to a common resolution;
\item increased accuracy of peak column densities of compact features,
due both to the resolution improvement and to taking proper account of 
temperature variations along the line of sight;
\item the temperature decomposition provides the potential ability to
distinguish different physical phenomena superposed along the line of sight,
as discussed above.
\end{enumerate}

We have processed all 163 tiles of the Hi-GAL survey, each of which
covers a $\sim2^\circ\times2^\circ$ region of the Galactic Plane.
The results, now publicly available online, consist of:
\begin{enumerate}
\item a total of 163 image cubes of differential column density with
6 arcsec spatial pixels and 12 values along the temperature axis, covering
the range 8--50 K in dust temperature;
\item corresponding image cubes of the uncertainties;
\item 2D maps of total column density and density-weighted mean dust
temperature, derived from the image cubes.
\end{enumerate}

In this paper we have presented some examples of analyses that can
be carried out with these data, focussing on the HII region bubble W5-E,
a star-forming filamentary complex in CMa OB1, and the radial distribution
of dust temperature in the Galaxy. In W5-E we have identified a 
strong temperature gradident in a pillar-like feature which may have 
implications for models of the bubble formation.
For CMa OB1 we confirm our previous findings regarding the 
temperature dependence of the form of the PDF of differential column
density, and also find evidence for spatial variations of that quantity
within the same star formation region.
Finally, we find that the radial dust 
temperature profile of the Galaxy shows a monotonic decrease from the
centre in rough agreement with published models of the interstellar radiation 
field, but that it also exhibits a prominent central plateau which is not 
predicted by such models and which therefore merits further investigation.

\section*{Acknowledgments}

We thank Thomas Haworth for an illuminating discussion, and Matt Griffin
for helpful comments on the manuscript. We also thank the referee for
helpful comments. We have made use of data from
the {\it Spitzer\/} Space Telescope, operated by the Jet Propulsion
Laboratory, California Institute of Technology, under contract with NASA.
This research has been supported by the EU-funded {\sc vialactea} Network (Ref. FP7-SPACE-607380).

\bsp

\label{lastpage}

\end{document}